# Diagonal Curvature in Second-Order Jahn–Teller Theory Can Be Negative: An Analytic Proof with First-Principles Confirmation in NH$_3$

.
Zhian Li,[1,3,*] Hanxiang Mi,[1,4,*] Xiyue Cheng,[1,2] Jürgen Köhler,[5] and Shuiquan Deng[1,2,†]

[1]State Key Laboratory of Functional Crystals and Devices, Fujian Institute of Research on the Structure of Matter (FJIRSM), Chinese Academy of Sciences (CAS), Fuzhou 350108, Fujian, P. R. China
[2]Fujian Science and Technology Innovation Laboratory for Optoelectronic Information of China, Fuzhou 350108 Fujian, P. R. China
[3]College of Chemistry and Materials Science, Fujian Normal University, Fuzhou 350117 Fujian, P. R. China
[4]University of Chinese Academy of Sciences, Beijing 100049, P. R. China
[5]Max Planck Institute for Solid State Research, Heisenbergstasse 1, 70569, Stuttgart, Germany

**ABSTRACT**: We demonstrate that the diagonal curvature in second-order Jahn–Teller theory (SOJTT) can be negative, invalidating its foundational positivity assumption. Using a frozen-phonon many-body expansion, we show no quantum-mechanical principle enforces its positivity; recasting into the Kohn–Sham framework further reveals that phonon-induced eigen-energy changes arise solely from electron–nuclear, Hartree, and exchange–correlation terms, with kinetic-energy contributions canceling identically. First-principles calculations on NH$_3$ place the D$_{3h}$ reference configuration at an A$_2''$ saddle point; Schwarz-inequality upper-bound analysis confines the HOMO–LUMO mixing contribution to ≤ 0.2 % of the total energy drop, while 99.8 % originates from the diagonal electron–nuclear term driving N (2s – 2p$_z$) electron redistribution. These findings overturn a core SOJTT tenet and establish that spontaneous structural symmetry breaking demands prior verification of a saddle point, not assumed positivity of the diagonal curvature.

The Second-Order Jahn-Teller (SOJT) effect, also known as Pseudo-Jahn-teller effect, has been widely used to explain diverse phenomena, including giant magnetoresistance, high-temperature superconductivity [1], chemical reaction dynamics [2] and structural distortions in molecules and solids [3-10]. Examples include distortions observed in graphene [7], ferroaxial-materials [10], and nonlinear optical crystals [11], among others. This approach now underpins explanations for EH$_3$ (E = N, P, As, or Sb) molecular geometries (via filled - empty orbital interactions) [12], acentric coordination of d$^0$ transition metal cations (e.g., Ti$^{4+}$, Nb$^{5+}$, W$^{6+}$, etc.) and ns$^2$ lone pair cations (e.g., Sn$^{2+}$, Se$^{4+}$, Te$^{4+}$, I$^{5+}$, etc.) [11], and SOJT-driven ferroaxial transitions in Na-based superionic conductors [10]. Even in α-PbO, initially interpreted through HOMO-LUMO interactions [13,14], later work reclassified its distortion as SOJT-mediated [8].

The SOJT effect was first proposed by Opik and Pryce (1957), who showed that non-degenerate electronic states can still induce structural instability [15]. Formally, SOJT effect arises from electron-phonon coupling treated within perturbation theory [16-19]. However, in practical applications to materials science, the theoretical prerequisites for perturbation expansions are seldom scrutinized. Consequently, SOJT theory (SOJTT) has evolved into a phenomenological descriptor for any system exhibiting HOMO-LUMO orbital mixing, irrespective of strict theoretical validity [8,11,20,21]. Critically, HOMO-LUMO mixing is not unique to SOJTT and may originate from other mechanisms. To conclusively establish SOJT-driven instability, one must rigorously validate the perturbation theory framework through quantitative comparison between the SOJTT predicted energy and first-principles computational results.

The SOJTT is usually formulated through a Hamiltonian, $\mathcal{H}$, expressed as the Taylor expansion [10,16,17,19,20,22]

$$\mathcal{H}(r, Q) = \mathcal{H}^{(0)}(r, 0) + \mathcal{H}^{(1)}Q + \frac{1}{2}\mathcal{H}^{(2)}Q^2 + o(Q^3), (1)$$

$$\mathcal{H}^{(1)} = \left(\frac{\partial \mathcal{H}(r,Q)}{\partial Q}\right)_{Q=0}, \mathcal{H}^{(2)} = \left(\frac{\partial^2 \mathcal{H}(r,Q)}{\partial Q^2}\right)_{Q=0}.$$

Here, $r$ is shorthand denoting the electronic degrees of freedom, and $Q$ denotes a specific phonon mode. In the frozen-phonon approach [23,24], $\mathcal{H}(r, Q)$ gives the total energy of the system with every atom fixed at the positions produced by a single phonon mode, the amplitude $Q$ of which is treated as a classical variable rather than as an operator. $\mathcal{H}^{(0)}$ denotes the Hamiltonian for the system with all atoms fixed at their "equilibrium" positions ($Q = 0$); the following terms are first- and second-order perturbations, and higher orders are dropped. Assume $\mathcal{H}^{(0)}$ is solvable, yielding eigenvalues $E^{(n)}$ and orthonormal eigenstates $|n\rangle$. Using these bases and second-order perturbation theory, the total energy is given by Eq. (2).

---

*These authors contributed equally to this work.
†Contact author: sdeng@fjirsm.ac.cn

$$E = E^{(0)} + \langle 0|\mathcal{H}^{(1)}|0\rangle Q + \frac{1}{2}\langle 0|\mathcal{H}^{(2)}|0\rangle Q^2 + \sum_{n\neq 0} \frac{|\langle 0|\mathcal{H}^{(1)}|n\rangle|^2}{E^{(0)} - E^{(n)}} Q^2 + \cdots \quad (2)$$

Where the first term, $E^{(0)}$, is the unperturbed ground-state energy; the second, $\langle 0|\mathcal{H}^{(1)}|0\rangle Q$, drives the first-order Jahn–Teller (FOJT) effect. If the electronic ground state $|0\rangle$ is degenerate at $Q = 0$, $\mathcal{H}^{(1)}$ has non-zero off-diagonal elements within that subspace, producing a linear energy term: $E(Q) \approx E^{(0)} + gQ + o(Q^2)$, $g \neq 0$. Consequently, $Q = 0$ is not a critical point of $E(Q)$, and the initial geometry is unstable—this is the FOJT effect. SOJT effect arises when FOJT effect is suppressed ($g = 0$); In SOJT studies the third term of Eq. (2) is customarily taken to be positive [10,16,17,19,20,22,25]. The usual justification is that this term depends only on the ground-state charge density, which favors the high-symmetry reference configuration [10,16,17,19,20,26]. Others trace the assumed sign to the repulsive character of the interaction [22,25]; either view casts the fourth term as decisive, since it must be sufficiently negative to outweigh the third and drive the distortion.

Two central challenges confronting SOJTT are whether the positive-definite character of $\langle 0|\mathcal{H}^{(2)}|0\rangle$ follows from any fundamental principle or is simply an assumption, and whether a qualitative indicator such as HOMO–LUMO mixing can be regarded as a reliable discriminator for the effect.

To address these challenges, we derive the explicit form of $\mathcal{H}^{(2)}$ in Eq. (2). In the Hamiltonian [27] given below (atomic units), only two terms depend on nuclear positions:

$$\mathcal{H}(\boldsymbol{r}, 0) = \sum_i -\frac{1}{2}\nabla_i^2 + \sum_{i>j} \frac{1}{|\boldsymbol{r}_i - \boldsymbol{r}_j|}$$

$$+ \sum_{i,I} \frac{-Z_I}{|\boldsymbol{r}_i - \boldsymbol{R}_I|} + \sum_{I>J} \frac{Z_I Z_J}{|\boldsymbol{R}_I - \boldsymbol{R}_J|}. \quad (3a)$$

$$H(\boldsymbol{r}, 0) = \sum_{I>J} \frac{Z_I Z_J}{|\boldsymbol{R}_I - \boldsymbol{R}_J|} + \sum_{i,I} \frac{-Z_I}{|\boldsymbol{r}_i - \boldsymbol{R}_I|}. \quad (3b)$$

Where $Z_I$ is the charge of the atomic core (nucleus plus frozen core electrons) centered at $\boldsymbol{R}_I$, and $\boldsymbol{r}_i$ is the position vector of $i$-th electron. If a phonon mode $Q$ displaces this core to $\boldsymbol{R}_I + \boldsymbol{u}_I$, Eq. (3b) becomes:

$$H(\boldsymbol{r}, Q) = \sum_{I>J} \frac{Z_I Z_J}{|\boldsymbol{R}_{IJ} - \boldsymbol{u}_{JI}|} + \sum_{i,I} \frac{-Z_I}{|\boldsymbol{R}_{Ii} - \boldsymbol{u}_I|}, \quad (4)$$

with $\boldsymbol{R}_{IJ} = \boldsymbol{R}_I - \boldsymbol{R}_J$, $\boldsymbol{u}_{JI} = \boldsymbol{u}_J - \boldsymbol{u}_I$, $\boldsymbol{R}_{Ii} = \boldsymbol{r}_i - \boldsymbol{R}_I$ and $|\boldsymbol{R}_{IJ}| \gg |\boldsymbol{u}_{JI}|$, $|\boldsymbol{R}_{Ii}| \gg |\boldsymbol{u}_I|$. The explicit second-order Hamiltonian of Eq. (2) is obtained only after Taylor-expanding Eq. (4) to second order in the nuclear displacements; the result is

$$\frac{1}{2}\mathcal{H}^{(2)}Q^2 = H_1^{(2)}(Q) + H_2^{(2)}(\boldsymbol{r}, Q), \quad (5)$$

with $H_1^{(2)}(Q)$ and $H_2^{(2)}(Q)$ given by

$$H_1^{(2)}(Q) = \sum_{I>J} \frac{Z_I Z_J}{2R_{IJ}^5}\left(-R_{IJ}^2 u_{JI}^2 + 3(\boldsymbol{R}_{IJ} \cdot \boldsymbol{u}_{JI})^2\right),$$

$$H_2^{(2)}(\boldsymbol{r}, Q) = \sum_{i,I} \frac{-Z_I}{2R_{Ii}^5}\left(-R_{Ii}^2 u_I^2 + 3(\boldsymbol{R}_{Ii} \cdot \boldsymbol{u}_I)^2\right).$$

Since $H_1^{(2)}(Q)$ does not depend on electronic coordinates, the expectation value $\langle 0|H_1^{(2)}(Q)|0\rangle$, corresponding to the electronic ground state at $Q = 0$, can be straightforwardly evaluated as follows:

$$\langle 0|H_1^{(2)}(Q)|0\rangle =$$

$$\begin{cases} 1/2 \sum_{I>J}(2 Z_I Z_J / R_{IJ}^3) u_{JI}^2 > 0 & (\boldsymbol{u}_{JI} \parallel \boldsymbol{R}_{IJ}, \text{longitudinal}), \\ -1/2 \sum_{I>J}(Z_I Z_J / R_{IJ}^3) \cdot u_{JI}^2 < 0 & (\boldsymbol{u}_{JI} \perp \boldsymbol{R}_{IJ}, \text{transverse}). \end{cases} \quad (6)$$

As evident from Eq. (6), the contribution of $H_1^{(2)}(Q)$ arising from the repulsive ion–ion term in Eq. (4) is positive for bond-stretching or compressive phonons, but negative for shear modes. And the positive contribution of the longitudinal mode is two times effective than the negative contribution of the transverse modes. For $H_2^{(2)}(\boldsymbol{r}, Q)$, longitudinal displacement ($\boldsymbol{u}_I \parallel \boldsymbol{R}_{Ii}$) gives $-1/2\sum_I(2Z_I/R_{Ii}^3)u_I^2 < 0$, whereas transverse displacement ($\boldsymbol{u}_I \perp \boldsymbol{R}_{Ii}$) yields $1/2\sum_I(Z_I/R_{Ii}^3) \cdot u_I^2 > 0$. Evaluated in the electronic ground state at $Q = 0$,

$$\langle 0|H_2^{(2)}(\boldsymbol{r}, Q)|0\rangle =$$

$$\begin{cases} -\frac{1}{2}\sum_{i,I}(2Z_I \int \frac{1}{R_{Ii}^3}\rho_0(\boldsymbol{r})d\boldsymbol{r}) u_I^2 < 0 & (\boldsymbol{u}_I \parallel \boldsymbol{R}_{Ii}, \text{longitudinal}), \\ \frac{1}{2}\sum_{i,I}(Z_I \int \frac{1}{R_{Ii}^3}\rho_0(\boldsymbol{r})d\boldsymbol{r}) u_I^2 > 0 & (\boldsymbol{u}_I \perp \boldsymbol{R}_{Ii}, \text{transverse}), \end{cases} \quad (7)$$

*These authors contributed equally to this work.
†Contact author: sdeng@fjirsm.ac.cn

where $\rho_0(\boldsymbol{r})$ is the ground-state electron density and the quantum expectation $\langle 0|1/R_{Ii}^3|0\rangle = \int \rho_0(\boldsymbol{r})\,1/R_{Ii}^3\,d\boldsymbol{r}$ is always positive. In contrast to $H_1^{(2)}(Q)$, the longitudinal mode now contributes negatively and is twice as effective as the positive contribution from the transverse mode. Because longitudinal nuclear motion modifies the electron–nuclear (e-N) distance ($\boldsymbol{R}_{Ii} \equiv \boldsymbol{r}_i - \boldsymbol{R}_I$), its contribution reflects the response of the ground-state radial electron distribution. Specifically, it entails the redistribution of electrons among orbitals with the same principal quantum number $n$ but different angular momenta $l$ (s, p, d, …). Transitions involving changes in $n$ are neglected here, as the electron–phonon interaction is intrinsically weak compared with other electronic interactions. On the other hand, the transverse displacement preserves the e–N distance; its contribution therefore reflects the response of the angular electron distribution, involving electron redistribution among orbitals with different magnetic quantum numbers $m$.

Eqs. (5-7) clearly demonstrate that $\langle 0|\mathcal{H}^{(2)}|0\rangle$ is positive-definite only if $\boldsymbol{u}_{JI} \parallel \boldsymbol{R}_{IJ}$ and $\boldsymbol{u}_I \perp \boldsymbol{R}_{Ii}$ are satisfied simultaneously. Since $\boldsymbol{R}_{Ii}$ is a vector operator, its expectation value is a single vector, so the above alignment conditions are almost impossible to enforce for every degree of freedom. Any deviation drives $\langle 0|\mathcal{H}^{(2)}|0\rangle$ negative or leaves its sign hostage to the phonon polarization and the initial structural details. These results show that the positivity of $\langle 0|\mathcal{H}^{(2)}|0\rangle$ is merely an assumption underlying the SOJTT framework. If one accepts this assumption, SOJTT predicts that the initial high-symmetry structure is unstable whenever the second-order state-mixing energy $\sum_{n\neq 0} \frac{|\langle 0|\mathcal{H}^{(1)}|n\rangle|^2}{E^{(0)} - E^{(n)}} Q^2$ dominates $\frac{1}{2}\langle 0|\mathcal{H}^{(2)}|0\rangle Q^2$, driving the system to a lower-energy distorted configuration. The ansatz demands that the direct product of the relevant electronic states and the phonon, $\Gamma^{(0)} \times \Gamma^{(n)} \times \Gamma^{(Q)}$, contain the totally symmetric irreducible representation; only then is the linear-coupling matrix element $\langle 0|\mathcal{H}^{(1)}|n\rangle$ non-vanishing [10,19]. In addition, the energy denominator $E^{(0)} - E^{(n)}$ must be sufficiently small for the negative state-mixing term to outweigh the positive $\frac{1}{2}\langle 0|\mathcal{H}^{(2)}|0\rangle Q^2$. These two conditions quantitatively invalidate the oversimplified HOMO–LUMO picture often invoked to rationalize SOJTT in practical cases.

The second-order contribution in Eq. (2) can be recast to a harmonic energy $\frac{1}{2}k_{tot}Q^2$, with the total stiffness

$$k_{tot} = \langle 0|\mathcal{H}^{(2)}|0\rangle + 2\sum_{n\neq 0}\frac{|\langle 0|\mathcal{H}^{(1)}|n\rangle|^2}{E^{(0)} - E^{(n)}}. \qquad (8)$$

The second term is always negative because the denominator $E^{(0)} - E^{(n)}$ is negative by definition. Within the adiabatic approximation [27] the total stiffness is the curvature of the ground-state energy surface, $\left.\partial^2 E/\partial Q^2\right|_{Q=0} = k_{tot}$. Stability of the reference configuration ($Q = 0$) therefore requires simultaneously $g \equiv \left.\partial E/\partial Q\right|_{Q=0} = 0$ and $k_{tot} > 0$. The first condition ($g = 0$) forces the initial geometry to sit at a critical point of $E(Q)$; the second ($k_{tot} > 0$) would make that point a local minimum. In the SOJTE regime $k_{tot}$ turns negative: the critical point is retained ($g = 0$) but the curvature is reversed, so the reference structure is unstable even when $\langle 0|\mathcal{H}^{(2)}|0\rangle$ is assumed positive. The $k_{tot} \leq 0$ sector subsumes the $\langle 0|\mathcal{H}^{(2)}|0\rangle \leq 0$ case that SOJTT explicitly discards. Here the adiabatic energy $E(Q)$ has vanishing gradient and non-positive curvature at $Q = 0$; the reference configuration therefore sits at a saddle point or maximum of the energy surface and is unstable against an infinitesimal symmetry-breaking distortion, triggering spontaneous symmetry breaking without external driving. This is qualitatively distinct from the $k_{tot} > 0$ scenario in which the system starts in a local minimum and symmetry breaking requires external control parameters (temperature, pressure, etc.) to invert the sign of $k_{tot}$ by reshuffling the diagonal and off-diagonal contributions in Eq. (8). Quantifying those two terms separately is essential for a rigorous test of the perturbative SOJTT ansatz.

However, solving Eq. (3) with true many-body wavefunctions remains prohibitively expensive; we therefore adopt a first-principles, density-functional framework [28,29] to obtain the required energies and matrix elements. Within DFT, the many-body Hamiltonian and its wave-functions (determinants or their linear combinations) are replaced by the Kohn–Sham (K-S) Hamiltonian and single-particle orbitals, slashing computational cost but adding an exchange-correlation (xc) term and redistributing weight among kinetic, e-N and Hartree contributions relative to the exact decomposition in Eq. (3). Accordingly, a theoretical bridge is required to relate the many-body formalism of Eqs. (1–8) to the DFT observables when tracing the driving force of a structural transition. Because $Q$ is not a fundamental variable in DFT, we carry out the expansion in terms of the charge-density response; the phonon coordinate enters only parametrically through the K–S Hamiltonian, wave functions, and their first- and second-order derivatives [30-32]. This yields the Kohn–Sham eigenvalue perturbation series:

$$\varepsilon_i = \varepsilon_i^0 + \langle\phi_i^0|V_{eff}^{(1)}|\phi_i^0\rangle + \langle\phi_i^0|V_{eff}^{(2)}|\phi_i^0\rangle + \sum_{m\neq i}\frac{|\langle\phi_m^0|V_{eff}^{(1)}|\phi_i^0\rangle|^2}{\varepsilon_i^0 - \varepsilon_m^0} + \cdots \qquad (9),$$

*These authors contributed equally to this work.
†Contact author: sdeng@fjirsm.ac.cn

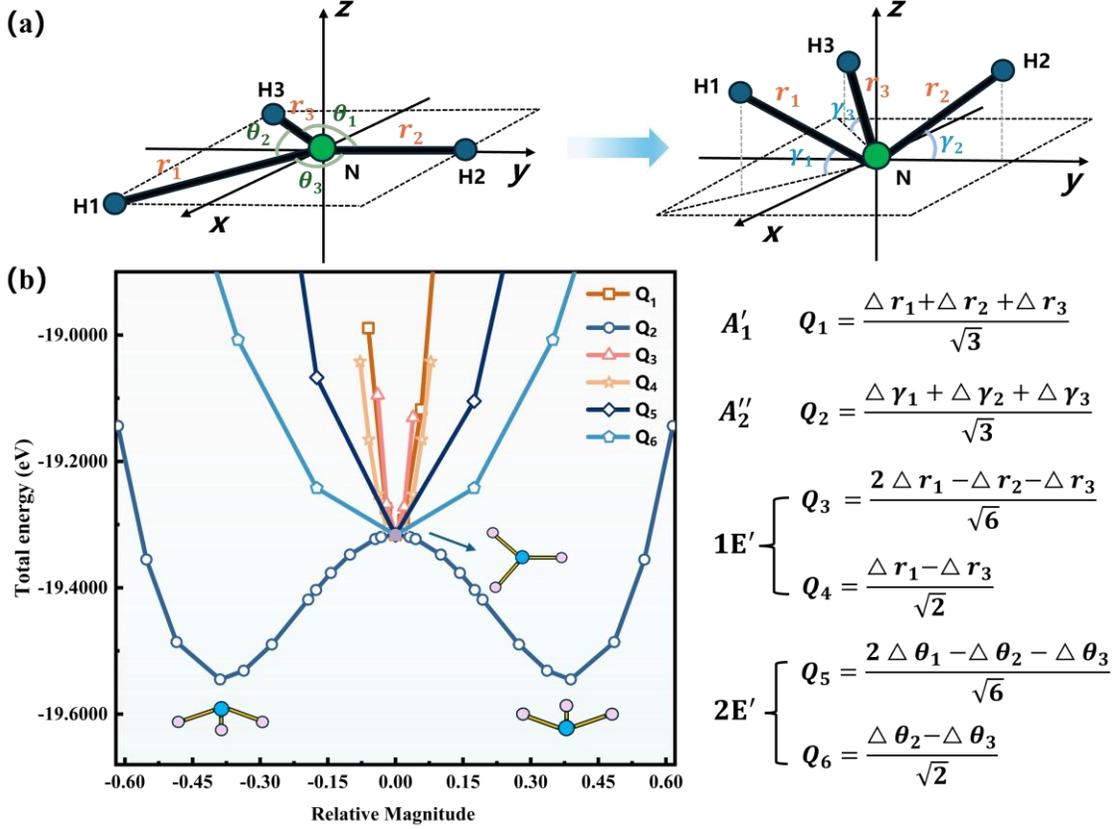

FIG 1. (a) Internal coordinates for the planar $D_{3h}$ reference (left) and experimental $C_{3v}$ pyramidal $NH_3$ (right); (b) Total-energy change versus amplitude for each of the six $D_{3h}$ phonon modes. Angle variations (in rad) and bond-length displacements are both expressed in the common unit u = 1.0214 Å, the experimental N–H distance.

where the effective potential is defined as $V_{eff} \equiv v_{ext}(r) + \int \rho(r')/|r-r'| dr' + v_{xc}(r)$ (see Supplemental Material (SM) for derivation details and further explanation) [33]. Equation (9) is formally identical to Eq. (2), but every quantity is now referenced to a single electron. The corresponding DFT total energy is obtained by summing the single-particle eigenvalues, subtracting the Hartree and exchange–correlation double-counting terms, and finally adding the Ewald ion–ion energy [34-39]. As Eq. (2) and the subsequent discussion show, the phonon-induced total-energy change has two microscopic sources: the e-N interaction and the Ewald ion–ion energy. In DFT the e-N variation is additionally mapped into two extra channels, the Hartree and the exchange–correlation (xc) energies, so the curvature acquires explicit Hartree and xc contributions that have no direct many-body counterpart. Equations (2) and (9) both show that the kinetic-energy change cancels identically, so any energy-decomposition scheme must be interpreted with care [34-38]. In practical SOJTT analyses attention is therefore restricted to the HOMO–LUMO piece of the last term in Eq. (9);

higher-lying states contribute negligibly because of the energy denominator [19].

We select $NH_3$, the canonical SOJTT paradigm [20], to interrogate the theory's foundational assumptions. First-principles computational details are provided in SM [33].

Fig. 1 shows the initial planar $D_{3h}$ $NH_3$ geometry together with the relaxed pyramidal $C_{3v}$ ground-state configuration; the six normal modes [40,41], expressed in internal coordinates, are also indicated. Using the first-principles method implemented in the VASP code [34,42], we traced the energy evolution from the initial hypothetical $D_{3h}$ geometry to the final actual $C_{3v}$ structure (see SM [33] for computational details). Fig. 1(b) shows the calculated total energy as a function of the amplitude for each of the six $D_{3h}$ normal modes. The initial planar structure sits at an energy minimum with respect to the $A_1'$ ($Q_1$), $1E'$ ($Q_3$, $Q_4$), and $2E'$ ($Q_5$, $Q_6$) modes, remaining stable against H-atom displacements induced by these symmetrized vibrations. In contrast, the planar structure is unstable against the $A_2''$ ($Q_2$) H-atom out-of-plane vibration, in accord with Pearson's group-theoretical analysis [20]. A cross-check with

*These authors contributed equally to this work.
†Contact author: sdeng@fjirsm.ac.cn

ABINIT [36,43] yields quantitatively consistent results (SM [33]). Scans along the $A_2''$ ($Q_2$) coordinate, generated by varying the phonon amplitude, yield potential-well depths of –0.2181 eV (LMTART [32,38]), –0.2169 eV (ABINIT) and –0.2180 eV (QE [44,45]), compared with –0.2286 eV (VASP [34,42]) (SM [33]). The close agreement, especially with the all-electron full-potential code LMTART, confirms that the initial $D_{3h}$ geometry sits at a saddle point of the total-energy surface $E(Q)$. The saddle-point instability at the $D_{3h}$ configuration, manifested as a negative curvature of the energy surface $E(Q)$ along the $A_2''$ phonon coordinate, provides the driving force for spontaneous symmetry breaking [46] to $C_{3v}$. The $A_2''$ phonon defines the reactive coordinate for the distortion, while the negative curvature supplies the thermodynamic impetus.

To quantify the off-diagonal contribution in Eq. (2) we estimate the matrix element $\langle 0|\mathcal{H}^{(1)}|n\rangle$ within the single-particle picture provided by DFT, i.e., $A_{HL}= \langle 0|\mathcal{H}^{(1)}|n\rangle \approx \langle \phi_{HOMO}^{(0)}|H^{(1)}|\phi_{LUMO}^{(0)}\rangle$ (see SM [33]). Here $\phi_i^{(0)}$ is the $i$-th Kohn–Sham eigenstate of $H(\mathbf{r}, Q = 0)$, and $H^{(1)} \equiv (\partial H(\mathbf{r},Q)/\partial Q)_{Q=0}$ is a DFT approximation of $\mathcal{H}^{(1)}$. Regarding $\phi_i^{(0)}$ as vectors in the Hilbert space $\mathcal{X}$, and noting that $H^{(1)}$ is Hermitian, Schwarz's inequality together with $\|\phi_{HOMO}^0\| = \|\phi_{LUMO}^0\| = 1$ gives $|A_{HL}| \leq \|H^{(1)}\|$. Here $\|H^{(1)}\|$ is the operator norm $\|H^{(1)}\| \equiv \sup_{\|v\|=1} \|H^{(1)}v\|$, $v \in \mathcal{X}$ [47]. Since $H^{(1)} \equiv (\partial H(\mathbf{r},Q)/\partial Q)_{Q=0}$, we approximate it numerically by the finite difference $H^{(1)} \approx \Delta H/\Delta Q$, where $\Delta H$ is the Kohn–Sham total energy change induced by an $A_2''$ phonon of small amplitude $\Delta Q$ in $NH_3$. Note that the $Q^2$ appearing in Eq. (2) is simply $(\Delta Q)^2$ relabeled under the SOJTT convention; the identical factor in the denominator of $H^{(1)} \approx \Delta H/\Delta Q$ therefore cancels. Within the frozen-phonon scheme we evaluate $\Delta H$, and the associated changes in the Kohn–Sham eigenvalues and eigenstates produced by an $A_2''$ distortion of amplitude $\Delta Q$. The calculated results for the $A_2''$ phonon with small displacements ($\Delta Q = 0.0318$) are listed in SM [33]. Because all eigenstates are normalized, the dominant eigenvalue shift is captured by the operator norm $\|H^{(1)}\|$, determined as described above. For the $A_2''$ phonon we obtain $\|H^{(1)}\| = 0.005541$ eV (evaluated with $v = \phi_{HOMO} \in \mathcal{X}$; the $Q$-factor cancels, as discussed above). Therefore $|A_{HL}| \leq 0.005541$ eV, and the leading contribution to $\sum_{n\neq 0} \frac{|\langle 0|\mathcal{H}^{(1)}|n\rangle|^2}{E^{(0)}-E^{(n)}} Q^2$ is at most $\sim 6.9\times 10^{-6}$ eV, merely $\sim 0.22\%$ of the total energy lowering ($\Delta H = 0.0032$ eV at $\Delta Q = 0.0318$). An estimate based on the Wolfsberg–Helmholtz relation [48] gives a vanishing contribution because $\phi_{HOMO}$ and $\phi_{LUMO}$ are orthogonal. Rotating LUMO orbital to $\phi_{LUMO}'$ introduces a finite overlap, but inserting the corresponding overlap integral and the frozen-phonon estimate ($\Delta H/\Delta Q$) into the numerator yields a contribution even smaller than the Schwarz upper bound (SM [33]). Hence the off-diagonal sum is negligible; the diagonal term alone is negative and fully accounts for the observed energy drop (Fig. 1 (b)).

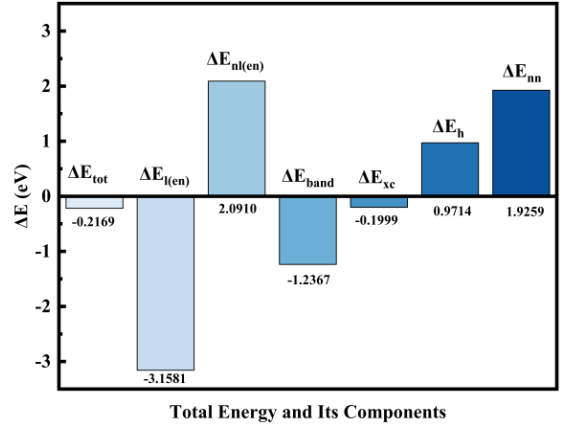

FIG 2. Total-energy and component-resolved changes from the planar $D_{3h}$ reference to the experimental $C_{3v}$ pyramidal $NH_3$ geometry.

Fig. 2 and SM [33] show the total and decomposed energy changes between the reference $D_{3h}$ and the relaxed $C_{3v}$ geometries of $NH_3$; the latter coincides with the experimentally established, realistic pyramidal structure. Full-potential (LMTART) and pseudopotential codes (VASP, ABINIT, QE) all yield the same total-energy drop, $\sim 0.22$ eV. Despite the different pseudopotentials and basis sets used in VASP, ABINIT, and QE, the variations in both the Ewald and Hartree contributions agree closely (SM [33]), confirming the consistency of the total-energy drop. Neither the Ewald nor the Hartree term favors the distortion. LMTART and VASP both find xc energy changes < 0.08 eV; ABINIT and QE yield somewhat larger xc stabilizations ($\sim 0.2$ eV), yet these remain far too small to offset the combined Ewald + Hartree increase. The consistent conclusion drawn from these data is that the distortion is driven by the e–N interaction, specifically its local component highlighted by the ABINIT decomposition (SM [33]), which contributes via the mechanism described by Eq. (7) and provides the negative curvature that converts the $D_{3h}$ saddle point into the $C_{3v}$ minimum.

ABINIT reproduces the stabilizing (negative) kinetic-energy trend reported by García-Fernández et al. [49], whereas LMTART gives the opposite sign and an absolute value roughly half as large (SM [33]). The perturbative treatment formally generates a phonon-induced change in kinetic energy, but this change is exactly cancelled when the eigenvalue is evaluated (see Eq. (9)).

*These authors contributed equally to this work.
†Contact author: sdeng@fjirsm.ac.cn

Guided by the state-mixing term in Eq. (2) or Eq. (9), bonding-change arguments have been advanced in the literature to rationalize distortions within SOJTT [10,19,22,25]. Our COHP analysis [50,51], however, shows a N–H bonding-energy loss of ~0.058 eV, i.e. an energy penalty that opposes the distortion (SM [33]). The calculations reveal that only the HOMO, LUMO, and the lowest $a_1'$ ($a_1$) orbital undergo appreciable changes (SM [33]). A significant fraction of N-2pz weight moves from the occupied HOMO into the unoccupied LUMO, while N-2s weight is transferred from the LUMO (and neighboring regions) into the HOMO. This s - p redistribution, driven by the mechanism in Eq. (7), differs from the HOMO–LUMO hybridization scenario invoked by SOJTT.

We have re-examined the central assumption of second-order Jahn–Teller theory with a frozen-phonon, many-body formulation and shown that the diagonal curvature is not necessarily positive, contrary to the theory's core premise. Recasting the many-body expansion into the Kohn–Sham framework, we show that the phonon-induced eigen-energy change is captured by the electron–nuclear, Hartree and exchange–correlation components, while the kinetic-energy variation cancels identically. First-principles DFT places the $D_{3h}$ NH$_3$ structure at a saddle point unstable exclusively against the $A_2''$ H-atom vibration. Schwarz-inequality and Wolfsberg–Helmholtz analyses confine the HOMO–LUMO state-mixing contribution to $\lesssim 0.2$ % of the total energy drop; the remaining 99.8 % originates from the diagonal e–N term that drives the N (2s–2p$_z$) electron redistribution, a conclusion cross-verified with full-potential and pseudopotential data. These findings overturn a foundational SOJTT premise and establish a microscopic origin for second-order, spontaneous-symmetry-breaking structural instabilities, underscoring that the reference structure must first be confirmed as a saddle point before invoking the theory.

## ACKNOWLEDGMENTS

This work was financially supported by the Strategic Priority Research Program of the Chinese Academy of Sciences (XDB1170000); National Natural Science Foundation (NSF) of China (62475265, 22031009); the Self-deployed Key Project of State Key Laboratory of Functional Crystals and Devices (GNJT-2025-ZD01). The computation facilities were provided by HZWTECH.

*These authors contributed equally to this work.
†Contact author: sdeng@fjirsm.ac.cn

[1] I. B. Bersuker, *The Jahn-Teller effect* (Cambridge University Press, Cambridge, 2006), p. 5-6.
[2] J. Rusho, J. Nichols, and J. Simons, Int. J. Quantum Chem. **48**, 309 (1993).
[3] P. Schwerdtfeger, P. D. W. Boyd, T. Fischer, P. Hunt, and M. Liddell, J. Am. Chem. Soc. **116**, 9620 (1994).
[4] H. Mizoguchi, P. M. Woodward, S. H. Byeon, and J. B. Parise, J. Am. Chem. Soc. **126**, 3175 (2004).
[5] B. E. F. Tiedemann and K. N. Raymond, Angew. Chem. **119**, 5064 (2007).
[6] Y. Inaguma, A. Aimi, Y. Shirako, D. Sakurai, D. Mori, H. Kojitani, M. Akaogi, and M. Nakayama, J. Am. Chem. Soc. **136**, 2748 (2014).
[7] C. Andreasen, T. Hao, J. Hatoum, and Z. M. Hossain, J. Appl. Phys. **130**, 034303 (2021).
[8] Y. Fu, S. Jin, and X. Y. Zhu, Nat. Rev. Chem **5**, 838 (2021).
[9] L. M. Sigmund, R. Maier, and L. Greb, Chem. Sci. **13**, 510 (2022).
[10] T. Nagai, Y. Mochizuki, S. Yoshida, and T. Kimura, J. Am. Chem. Soc. **145**, 8090 (2023).
[11] P. S. Halasyamani and K. R. Poeppelmeier, Chem. Mater. **10**, 2753 (1998).
[12] H. B. Wedler, P. Wendelboe, D. J. Tantillo, and P. P. Power, Dalton Trans. **49**, 5175 (2020).
[13] G. W. Watson and S. C. Parker, Phys. Rev. B **59**, 8481 (1999).
[14] A. Walsh and G. W. Watson, J. Solid State Chem. **178**, 1422 (2005).
[15] U. Opik and M. H. L. Pryce, Proc. R. Soc. A **238**, 425 (1957).
[16] R. F. W. Bader, Mol. Phys. **3**, 137 (1960).
[17] R. F. W. Bader, Can. J. Chem **40**, 1164 (1962).
[18] I. B. Bersuker, *The Jahn-Teller effect* (Cambridge University Press, Cambridge, 2006), p. 159-160.
[19] S. Yoshida, H. Akamatsu, and K. Hayashi, Phys. Rev. Lett. **127**, 215701 (2021).
[20] R. G. Pearson, J. Mol. Struct. **103**, 25 (1983).
[21] X. D. Huai and T. T. Tran, Annu. Rev. Mater. Res. **53**, 253 (2023).
[22] D. Hickox-Young, D. Puggioni, and J. M. Rondinelli, Phys. Rev. B **102**, 014108 (2020).
[23] H. Wendel and R. M. Martin, Phys. Rev. B **19**, 5251 (1979).
[24] J. Xu, X. Huang, X. Cheng, M. H. Whangbo, and S. Deng, Angew. Chem. Int. Edit. **61**, e2021116404 (2022).
[25] J. M. Rondinelli, A. S. Eidelson, and N. A. Spaldin, Phys. Rev. B **79**, 205119 (2009).
[26] R. G. Pearson, J. Am. Chem. Soc. **91**, 4947 (1969).

[27] G. Grimvall, *The electron-phonon interaction in metals.* (North-Holland Publishing Company, New York, 1981), p. 36-42.
[28] P. Hohenberg and W. Kohn, Phys. Rev. B **136**, B864 (1964).
[29] W. Kohn and L. J. Sham, Phys. Rev. **140**, A1133 (1965).
[30] S. Baroni, P. Giannozzi, and A. Testa, Phys. Rev. Lett. **58**, 1861 (1987).
[31] X. Gonze, Phys. Rev. A **52**, 1096 (1995).
[32] S. Y. Savrasov, Phys. Rev. B **54**, 16470 (1996).
[33] See Supplemental Material at http://link.aps.org/supplemental/10.1103/PhysRevLett.xxxxx for detailed derivations of the perturbative treatment of the Kohn–Sham equations, computational details, total energy decomposition, eigenvalue variations, contributions of off-diagonal terms to the second-order perturbation energy, bonding-energy analysis, and molecular orbital analysis.
[34] G. Kresse and J. Furthmuller, Phys. Rev. B **54**, 11169 (1996).
[35] G. Kresse and D. Joubert, Phys. Rev. B **59**, 1758 (1999).
[36] X. Gonze *et al.*, Comput. Phys. Commun. **180**, 2582 (2009).
[37] S. Poncé, G. Antonius, P. Boulanger, E. Cannuccia, A. Marini, M. Côté, and X. Gonze, Comput. Mater. Sci. **83**, 341 (2014).
[38] S. Y. Savrasov and D. Y. Savrasov, Phys. Rev. B **46**, 12181 (1992).
[39] S. Lundqvist and N. H. Norman, *Theory of the Inhomogeneous Electron Gas* (Plenum Press, New York, 2013).
[40] M. S. Schuurman, W. D. Allen, and H. F. Schaefer, J. Comput. Chem. **26**, 1106 (2005).
[41] C. R. Rodenas, J. V. Quesada, E. M. Torres, and J. J. Gonzalez, J. Mol. Model. **20**, 2222 (2014).
[42] G. Kresse and J. Furthmüller, Comput. Mater. Sci. **6**, 15 (1996).
[43] X. Gonze *et al.*, Comput. Phys. Commun. **205**, 106 (2016).
[44] P. Giannozzi *et al.*, J. Phys. Condens. Matter **29**, 465901 (2017).
[45] P. Giannozzi *et al.*, J. Chem. Phys. **152**, 154105 (2020).
[46] P. W. Anderson, *Basic Notions of Condensed Matter Physics* (Benjamin Cummings, Menlo Park, CA, 1984), p. 8-11.
[47] E. Zeidler, *Applied functional analysis: applications to mathematical physics.* (Springer, New York, 1995), p. 70.
[48] M. Wolfsberg and L. Helmholz, J. Chem. Phys. **20**, 837 (1952).
[49] P. Garcia-Fernandez, J. A. Aramburu, M. Moreno, M. Zlatar, and M. Gruden-Pavlovic, J. Chem. Theory Comput. **10**, 1824 (2014).
[50] R. Dronskowski and P. E. Blochl, J. Phys. Chem. **97**, 8617 (1993).
[51] S. Maintz, V. L. Deringer, A. L. Tchougreeff, and R. Dronskowski, J. Comput. Chem. **37**, 1030 (2016).

*These authors contributed equally to this work.
†Contact author: sdeng@fjirsm.ac.cn

# Supporting Material For

# Diagonal Curvature in Second-Order Jahn–Teller Theory Can Be Negative: An Analytic Proof with First-Principles Confirmation in NH$_3$

Zhian Li,[1,3,*] Hanxiang Mi,[1,4,*] Xiyue Cheng,[1,2] Jürgen Köhler,[5] and Shuiquan Deng[1,2,†]

[1]*State Key Laboratory of Functional Crystals and Devices, Fujian Institute of Research on the Structure of Matter (FJIRSM), Chinese Academy of Sciences (CAS), Fuzhou 350108, Fujian, P. R. China*
[2]*Fujian Science and Technology Innovation Laboratory for Optoelectronic Information of China, Fuzhou 350108 Fujian, P. R. China*
[3]*College of Chemistry and Materials Science, Fujian Normal University, Fuzhou 350117 Fujian, P. R. China*
[4]*University of Chinese Academy of Sciences, Beijing 100049, P. R. China*
[5]*Max Planck Institute for Solid State Research, Heisenbergstasse 1, 70569, Stuttgart, Germany*

# Contents

# S-I. PERTURBATION TREATMENT OF KOHN-SHAM EQUATION

Traditional second-order Jahn–Teller theory is formulated within a many-body framework whose computational demands have limited quantitative applications. We therefore adopt the Kohn–Sham perturbation expansion pioneered by Gonze [1] and developed by Savrasov [2] et al.[3]. for linear-response phonon calculations, but focus on second-order terms at saddle-point geometries rather than equilibrium, allowing direct evaluation of the curvature that drives spontaneous symmetry breaking.

$$\left[-\frac{1}{2}\nabla^2 + v_{\text{eff}}(\mathbf{r})\right]\psi_i(\mathbf{r}) = \varepsilon_i \cdot \psi_i(\mathbf{r}) \ldots(S1) \text{ Kohn-Sham Equation}$$

where the subscript labels the energy level (molecule) or band index (solid), and the effective potential is defined as

$$v_{\text{eff}}(\mathbf{r}) = v_{\text{ext}}(\mathbf{r}) + \int \frac{\rho(\mathbf{r}')}{|\mathbf{r}-\mathbf{r}'|}d\mathbf{r}' + v_{\text{xc}}(\mathbf{r}') \ldots(S2).$$

The external potential $v_{\text{ext}}(\mathbf{r})$ is defined as

$$v_{\text{ext}}(\mathbf{r}) := -\sum_{I,t}\frac{Z_I e^2}{\left|\mathbf{r}-\mathbf{R}_I-\mathbf{t}-\Delta R_{I,t}^{qj}\right|} \ldots(S3).$$

In Eq. S3, the displacement of atom $I$ in the unit cell whose origin is translated by the lattice vector $\mathbf{t}$ is written as

$$\Delta R_{I,t}^{qj} = Q_{I,t}^{qj}\exp(i\mathbf{q}\cdot\mathbf{t}) + Q_{I,t}^{qj*}\exp(-i\mathbf{q}\cdot\mathbf{t})\ldots(S4).$$

Where $\mathbf{q}j$ labels a phonon mode and $\mathbf{R}_I$ is the equilibrium position of atom $I$ in the unit cell (or molecule).

$$Q_{I,t}^{qj} = \frac{A_{qj}}{\sqrt{m_I}}e_I^{qj}\exp(i\mathbf{q}\cdot\mathbf{R}_I)\ldots(S5)$$

In Eq. (S5), $A_{qj}$ is the vibrational amplitude, $m_I$ the mass of atom $I$, and $e_I^{qj}$ the polarization vector of phonon $\mathbf{q}j$. Equations (S3)–(S5) form the basis for linear-response calculations of phonon spectra [4] in real materials. Owing to $\left|\Delta R_{I,t}^{qj}\right| \ll |\mathbf{r}-\mathbf{R}_I-\mathbf{t}|$, the perturbation expansion of $v_{\text{ext}}(\mathbf{r})$ converges uniformly and yields a $Q$-ordered series identical in structure to the second term of Eq. (4) in the main text. Since our goal is to trace how the many-body energy variation is mapped onto the Kohn–Sham form, we do not need an explicit expansion in the individual normal-mode coordinates $Q_{I,t}^{qj}$. For the molecule we set $\mathbf{q} = 0$ and $\mathbf{t} = 0$, and regard $Q_I^j$ no longer as a phonon amplitude but as a continuous distortion coordinate whose polarization is fixed by the molecular point group (see Fig. 1 of the main text).

$$v_{\text{ext}}(r) = v_{\text{ext}}^{(0)}(r) + \lambda v_{\text{ext}}^{(1)}(r) + \lambda^2 v_{\text{ext}}^{(2)}(r) + \mathcal{O}(\lambda^2)\ldots\ldots(S6)$$

$$\rho(r) = \rho^{(0)}(r) + \lambda\rho^{(1)}(r) + \lambda^2\rho^{(2)}(r) + \mathcal{O}(\lambda^2)\ldots\ldots(S7a)$$

$$\rho^{(0)}(r) = \sum_i \psi_i^{(0)*}(r)\psi_i^{(0)}(r)\ldots\ldots(S7b)$$

$$\rho^{(1)}(r) = \sum_i \left[\psi_i^{(0)*}(r)\psi_i^{(1)}(r) + \psi_i^{(1)*}(r)\psi_i^{(0)}(r)\right]\ldots\ldots(S7b)$$

$$\rho^{(2)}(r) = \sum_i \left[\psi_i^{(0)*}(r)\psi_i^{(2)}(r) + \psi_i^{(1)*}(r)\psi_i^{(1)}(r) + \psi_i^{(2)*}(r)\psi_i^{(0)}(r)\right]\ldots\ldots(S7c)$$

$$\psi_i(r) = \psi_i^{(0)}(r) + \lambda\psi_i^{(1)}(r) + \lambda^2\psi_i^{(2)}(r) + \mathcal{O}(\lambda^2)\ldots\ldots(S8)$$

$$v_{\text{xc}}(r) = v_{\text{xc}}^{(0)}(r) + \lambda v_{\text{xc}}^{(1)}(r) + \lambda^2 v_{\text{xc}}^{(2)}(r) + \mathcal{O}(\lambda^2)\ldots\ldots(S9)$$

$$\varepsilon_i = \varepsilon_i^{(0)} + \lambda\varepsilon_i^{(1)} + \lambda^2\varepsilon_i^{(2)} + \mathcal{O}(\lambda^2)\ldots\ldots(S10)$$

Assuming the orthonormality of the zero-order wave functions and the Hermitian property of the zero-order Hamiltonian, $\widehat{H}^0 = -\frac{1}{2}\nabla^2 + v_{\text{eff}}^0$, one immediately obtains:

$$\lambda^0: \quad \widehat{H}^0 \psi_i^0 = \varepsilon_i^0 \psi_i^0 \ldots\ldots(S11)$$

$$\lambda^1: \quad \widehat{H}^0 \psi_i^{(1)}(r) + v_{\text{eff}}^{(1)} \psi_i^{(0)}(r) = \varepsilon_i^{(0)} \psi_i^{(1)}(r) + \varepsilon_i^{(1)} \psi_i^{(0)}$$

$$|\psi_i^{(1)}\rangle = \sum_{m \neq i} \frac{\left|\langle\psi_m^0|v_{\text{eff}}^{(1)}|\psi_i^0\rangle\right|^2}{\varepsilon_i^0 - \varepsilon_m^0} \cdot |\psi_m^0\rangle$$

$$\varepsilon_i^{(1)} = \langle\psi_i^0|v_{\text{eff}}^{(1)}|\psi_i^0\rangle\ldots\ldots(S12)$$

$$\lambda^2: \quad \widehat{H}^0|\psi_i^{(2)}\rangle + v_{\text{eff}}^{(1)}|\psi_i^{(1)}\rangle + v_{\text{eff}}^{(2)}|\psi_i^0\rangle = \varepsilon_i^{(0)}|\psi_i^{(2)}\rangle + \varepsilon_i^{(1)}|\psi_i^{(1)}\rangle + \varepsilon_i^{(2)}|\psi_i^0\rangle$$

$$\varepsilon_i^{(2)} = \langle\psi_i^0|v_{\text{eff}}^{(1)}|\psi_i^{(1)}\rangle + \langle\psi_i^0|v_{\text{eff}}^{(2)}|\psi_i^0\rangle$$

$$\varepsilon_i^{(2)} = \langle\psi_i^0|v_{\text{eff}}^{(2)}|\psi_i^0\rangle + \sum_{m \neq i} \frac{\left|\langle\psi_m^0|v_{\text{eff}}^{(1)}|\psi_i^0\rangle\right|^2}{\varepsilon_i^0 - \varepsilon_m^0}\ldots\ldots(S13)$$

Using Eqs. (S10, S12, S13), one easily obtains Eq. (S14), which is the Eq. (9) in our main text.

$$\varepsilon_i = \varepsilon_i^0 + \langle\phi_i^0|v_{eff}^{(1)}|\phi_i^0\rangle + \langle\phi_i^0|v_{eff}^{(2)}|\phi_i^0\rangle + \sum_{m \neq i} \frac{\left|\langle\phi_m^0|v_{eff}^{(1)}|\phi_i^0\rangle\right|^2}{\varepsilon_i^0 - \varepsilon_m^0} + \cdots \quad (S14).$$

Equations (S13)–(S14) show that the kinetic-energy change makes no net contribution to the phonon-induced variation of a Kohn–Sham eigenvalue; it is exactly cancelled, even though the kinetic energy itself is modified by the distortion. Equations (S2) and (S14) indicate clearly that the distortion energy partition into three Kohn–Sham channels—electron–nuclear, Hartree and exchange–correlation—while the many-body formalism routes it solely through the electron–nuclear term (plus the Ewald contribution common to both). Because code-dependent decomposition schemes differ [5-9], the true driving force must be identified with care.

## S-II. COMPUTATIONAL DETAILS

In this work the electron–phonon system is handled within the same adiabatic (Born–Oppenheimer) [10] framework adopted by second-order Jahn–Teller theory: the electronic problem is solved for fixed nuclear positions, while the phonons are treated classically via the frozen-phonon method [11,12], yielding an effective potential that parametrically depends on the vibrational coordinate $Q$. This ansatz is implemented at the density-functional level, reducing the many-body Hamiltonian of Eqs. (1) – (3) to an effective single-particle problem [13,14]. Within this picture the electronic ground-state density is constructed from the occupied Kohn–Sham orbitals, and the total energy is obtained from the corresponding single-particle eigenvalues supplemented by nuclear–nuclear repulsion and double-counting corrections [6,7,9]; correlation beyond the local density approximation (LDA) exchange–correlation functional is therefore approximate [15]. Because $\varepsilon_i^0 - \varepsilon_m^0$ in Eq. (S14) grows rapidly for deep-occupied and high-unoccupied states, replacing the many-body gap $E^{(0)} - E^{(n)}$ in Eq. (8) of the main text with the Kohn–Sham HOMO–LUMO difference $\varepsilon_{HOMO} - \varepsilon_{LUMO}$ is a standard and acceptable approximation [16-18].

### A. VASP calculations and structural models

The Perdew-Burke-Ernzerhof (PBE) functional within the Generalized Gradient Approximation (GGA) [4] was employed to account for exchange-correlation effects. The projector augmented wave (PAW) method [19] was utilized to describe the valence electron wavefunctions. The pseudopotentials treated the $2s^22p^3$ and $1s^1$ electrons as valence for N and H, respectively. A 2×2×2 k-point mesh was applied for Brillouin zone sampling. A plane-wave basis set with an energy cutoff of 600 eV was adopted to ensure the convergence of both energy differences (1.0 ×10⁻⁸ eV) and stress tensors (0.1 meV/ Å). The tetrahedron method incorporating Blöchl corrections was applied to integrations within the first Brillouin zone. We analyzed the bonding characteristics using the Crystal Orbital Hamilton Population (COHP) method with the LOBSTER package [20-22].

To minimize intermolecular interactions, a large cubic simulation cell (a = b = c = 10.00 Å) was constructed. The initial $NH_3$ configuration, sourced from the ChemSpider database, was first structurally optimized in VASP to determine its theoretical $C_{3v}$ symmetric equilibrium geometry. The N atom of this optimized molecule was then placed at the center of the simulation cell for subsequent calculations. Using the optimized bond lengths, an initial planar $D_{3h}$ reference geometry for $NH_3$ was constructed.

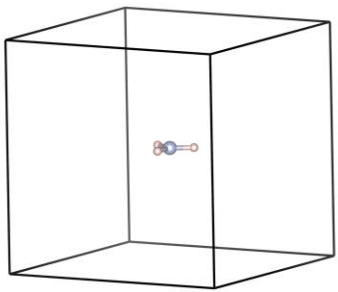

FIG. S1. $NH_3$ in its planar $D_{3h}$ reference configuration, positioned at the center of the cubic simulation cell.

### B. Quantum ESPRESSO (QE) calculations

The calculations were carried out using the Quantum ESPRESSO (QE) package [23-25] with optimized norm-conserving Vanderbilt (ONCV) pseudopotentials [26]. A plane-wave cutoff energy of 60 Ry was employed, and Brillouin-zone sampling was restricted to the $\Gamma$ point (1 × 1 × 1 k-point mesh). The self-consistent field (SCF) energy was converged to $1.0 \times 10^{-8}$ Ry.

### C. FP-LMTO calculations

Electronic structure calculations were performed using the LMTART code [2] within the full-potential linearized muffin-tin orbital (FP-LMTO) framework. The LMTO basis was constructed in the Bare representation, and the full-potential treatment was implemented via the PLW approach. Exchange–correlation effects were accounted for using the Vosko LDA. The valence states of nitrogen and hydrogen were represented by LMTO basis functions constructed from the 2s/2p and 1s/2p orbitals, respectively. Within the muffin-tin spheres, both wavefunctions and charge densities were expanded in spherical harmonics up to a maximum angular momentum of $l$max =

6. In the interstitial region, the charge density and potential were expressed as plane-wave expansions. The plane-wave cutoff energy, automatically determined by the code, reached approximately $9.36 \times 10^2$ Ry, corresponding to $4.08 \times 10^5$ plane waves in the present calculations. Self-consistency was achieved at the Γ-point alone, with an energy convergence criterion of $1.0 \times 10^{-7}$ Ry.

### D. ABINIT calculations

Calculations using the ABINIT package [7,27] employed the ONCV pseudopotentials [26]. With the following parameters: a plane-wave cutoff energy of 60 Ha, a 1×1×1 k-point mesh, and an SCF energy convergence threshold of $1.0 \times 10^{-10}$ Hartree. Figure S2 shows the calculated total energy as a function of the amplitude for each of the six $D_{3h}$ normal modes.

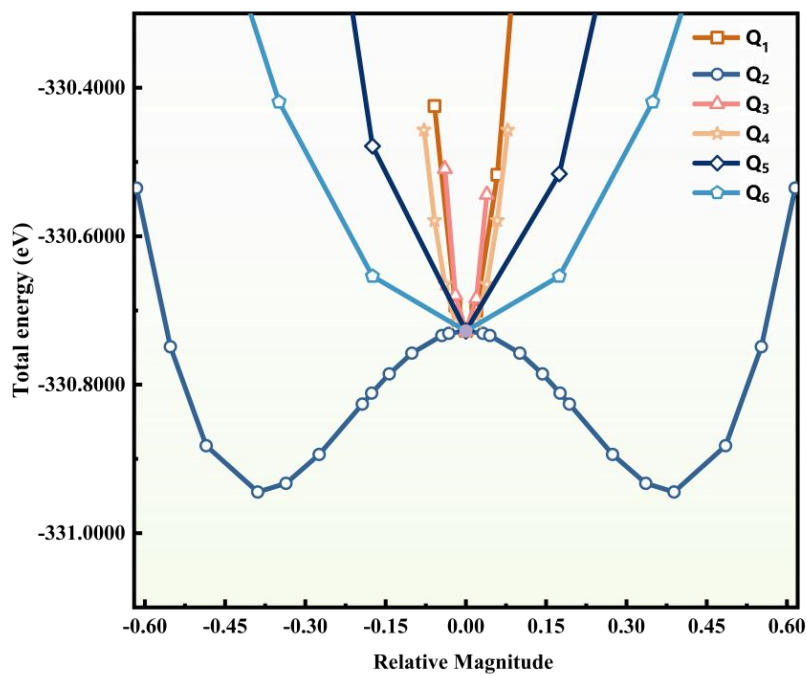

FIG. S2. Variation of the total energy with respect to the vibrational amplitude for all six phonon modes of $D_{3h}$-$NH_3$, as obtained from ABINIT calculations.

# S-III. TOTAL ENERGY AND COMPONENT ANALYSIS FROM LMTART, ABINIT, QE AND VASP CALCULATIONS

TABLE S1. LMTART-calculated variations in the total energy and its components between the reference $NH_3$ geometry ($\Delta Q = 0$, $D_{3h}$) and the relaxed structure ($\Delta Q = 0.3889$ [u], $C_{3v}$).

| $NH_3$ | reference structure $E_{ref}$ (eV) | relaxed structure $E_{rel}$ (eV) | $\Delta E = (E_{rel} - E_{ref})$ (eV) |
|---|---|---|---|
| $E_{coulomb}$ | -2852.8348 | -2853.8950 | -1.0602 |
| $E_{exch-corr}$ | -203.1238 | -203.0959 | 0.0278 |
| $E_{Band\ energy}$ | 172.9328 | 172.6461 | -0.2867 |
| $E_{tot}$ | -1523.6885 | -1523.9066 | -0.2181 |

Note: The magnitude of $\Delta Q$ for all phonons is in units of the N–H bond length, [u] = 1.0214 Å.

TABLE S2. ABINIT-calculated variations in the total energy and its components between the reference $NH_3$ geometry ($\Delta Q = 0$, $D_{3h}$) and the relaxed structure ($\Delta Q = 0.3889$ [u], $C_{3v}$).

| $NH_3$ | reference structure $E_{ref}$ (eV) | relaxed structure $E_{rel}$ (eV) | $\Delta E = (E_{rel} - E_{ref})$ (eV) |
|---|---|---|---|
| $E_{nn}$ | 105.9107 | 107.8366 | 1.9259 |
| $E_h$ | 356.6247 | 357.5961 | 0.9714 |
| $E_{xc}$ | -111.8810 | -112.0809 | -0.1999 |
| $E_k$ | 242.8245 | 240.9772 | -1.8473 |
| $E_{Band\ energy}$ | -97.4901 | -98.7268 | -1.2367 |
| $E_{nl}\ (E_{en})$ | -39.0047 | -36.9137 | 2.0910 |
| $E_l\ (E_{en})$ | -885.4050 | -888.5631 | -3.1581 |
| $E_{psp}\ (E_{en})$ | 0.2031 | 0.2031 | 0.0000 |
| $E_{tot}$ | -330.7277 | -330.9446 | -0.2169 |

TABLE S3. QE-calculated variations in the total energy and its components between the reference NH$_3$ geometry ($\Delta Q = 0$, D$_{3h}$) and the relaxed structure ($\Delta Q = 0.3889$ [u], C$_{3v}$).

| NH$_3$ | reference structure $E_{ref}$ (eV) | relaxed structure $E_{rel}$ (eV) | $\Delta E = (E_{rel} - E_{ref})$ (eV) |
|---|---|---|---|
| $E_{nn}$ | 105.9107 | 107.8366 | 1.9259 |
| $E_h$ | 356.5616 | 357.5267 | 0.9650 |
| $E_{xc}$ | -111.8476 | -112.0474 | -0.1998 |
| one-electron contribution | -681.2768 | -684.1861 | -2.9092 |
| $E_{tot}$ | -330.6521 | -330.8702 | -0.2180 |

TABLE S4. VASP-calculated variations in the total energy and its components between the reference NH$_3$ geometry ($\Delta Q = 0$, D$_{3h}$) and the relaxed structure ($\Delta Q = 0.3889$ [u], C$_{3v}$).

| NH$_3$ | reference structure $E_{ref}$ (eV) | relaxed structure $E_{rel}$ (eV) | $\Delta E = (E_{rel} - E_{ref})$ (eV) |
|---|---|---|---|
| Ewald energy | 105.9117 | 107.8377 | 1.9260 |
| -Hartree energy | -357.1436 | -358.2235 | -1.0781 |
| -V(xc)+Exc | 28.8042 | 28.8714 | 0.0671 |
| -exchange | 0.0000 | 0.0000 | 0.0000 |
| eigenvalues EBANDS | -97.5149 | -98.7305 | -1.2173 |
| alpha Z | 0.2309 | 0.2309 | 0.0000 |
| atomic energy EATOM | 302.0023 | 302.0023 | 0.0000 |
| PAW double counting | -1.6072 | -1.5333 | 0.0739 |
| $E_{tot}$ | -19.3165 | -19.5451 | -0.2286 |

TABLE S5. VASP-calculated variations in the total energy and its components between the reference $NH_3$ geometry ($\Delta Q = 0$, $D_{3h}$) and the distorted structure with small phonon amplitude ($\Delta Q = 0.0318$ [u], $C_{3v}$).

| $NH_3$ | reference structure $E_{ref}$ (eV) | distorted structure $E_{dis}$ (eV) | $\Delta E = (E_{dis} - E_{ref})$ (eV) |
|---|---|---|---|
| Ewald energy | 105.9117 | 105.9240 | 0.0123 |
| -Hartree energy | -357.1441 | -357.1498 | -0.0057 |
| -V(xc)+Exc | 28.8042 | 28.8048 | 0.0006 |
| -exchange | 0.0000 | 0.0000 | 0.0000 |
| eigenvalues EBANDS | -97.5144 | -97.5253 | -0.0109 |
| alpha Z | 0.2309 | 0.2309 | 0.0000 |
| atomic energy EATOM | 302.0023 | 302.0023 | 0.0000 |
| PAW double counting | -1.6072 | -1.6065 | 0.0007 |
| $E_{tot}$ | -19.3165 | -19.3197 | -0.0032 |

TABLE S6 ABINIT-calculated variations in the total energy and its components between the reference $NH_3$ geometry ($\Delta Q = 0$, $D_{3h}$) and the distorted structure with small phonon amplitude ($\Delta Q = 0.0318$ [u], $C_{3v}$).

| $NH_3$ | reference structure $E_{ref}$ (eV) | distorted structure $E_{dis}$ (eV) | $\Delta E = (E_{dis} - E_{ref})$ (eV) |
|---|---|---|---|
| $E_{nn}$ | 105.9107 | 105.9229 | 0.0122 |
| $E_h$ | 356.6247 | 356.6302 | 0.0055 |
| $E_{xc}$ | -111.8810 | -111.8828 | -0.0018 |
| $E_k$ | 242.8245 | 242.8074 | -0.0171 |
| $E_{Band\ energy}$ | -97.4901 | -97.5007 | -0.0106 |
| $E_{nl}$ ($E_{en}$) | -39.0047 | -38.9866 | 0.0181 |
| $E_l$ ($E_{en}$) | -885.4050 | -885.4251 | -0.0201 |
| $E_{psp}$ ($E_{en}$) | 0.2031 | 0.2031 | 0.0000 |
| $E_{tot}$ | -330.7277 | -330.7308 | -0.0031 |

# S-IV. CALCULATED VARIATIONS IN THE ENERGY EIGENVALUES.

TABLE S7. VASP-calculated variations in the energy eigenvalues between the reference $NH_3$ geometry ($\Delta Q = 0$, $D_{3h}$) and the distorted structure with small phonon amplitude ($\Delta Q = 0.0318$ [u], $C_{3v}$).

| Energy levels (n) | Reference struct. (Irr.: $\varepsilon_n^R$) (eV) | Distorted struct. (Irr.: $\varepsilon_n^D$) (eV) | Variation of eigenvalue (eV) ($\varepsilon_n^D - \varepsilon_n^R$) |
|---|---|---|---|
| 7 | 2e': 0.704977 | 2e: 0.705110 | 0.000132 |
| 6 | 2e': 0.320692 | 2e: 0.320708 | 0.000016 |
| 5 | $2a_1$': -0.591680 | $3a_1$: -0.591957 | -0.000277 |
| 4 | $a_2$": -5.230485 | $2a_1$: -5.236026 | -0.005541 |
| 3 | e': -11.540474 | e: -11.538046 | 0.002429 |
| 2 | e': -11.540745 | e: -11.538317 | 0.002428 |
| 1 | $a_1$': -20.445757 | $a_1$: = -20.450263 | -0.004506 |

Note: Irr. Irreducible representation

# S-V. CONTRIBUTIONS OF OFF-DIAGONAL TERM TO THE SECOND-ORDER PERTURBATION ENERGY, ESTIMATED VIA THE WOLFSBERG–HELMHOLTZ TYPE METHOD. (VASP)

$\Delta E = (-19.3197)$ eV $- (-19.3165)$ eV $= -0.0032$ eV. Total energy change

$\Delta Q = 0.0318$ [u]. ([u] =1.0214 Å). Magnitude of the distortion phonon mode ($A_2''$)

$\langle HOMO \mid LUMO' \rangle = 0.28$, $LUMO'$ is the LUMO of the distorted structure.

$$\sum_{n\neq 0} \frac{|\langle 0|\mathcal{H}^{(1)}|n\rangle|^2}{E^{(0)} - E^{(n)}} Q^2 \approx \frac{\left(\frac{\partial \mathcal{H}(r,Q)}{\partial Q}\right)^2_{Q=0} \langle HOMO \mid LUMO' \rangle^2 \Delta Q^2}{E^{(0)} - E^{(1)}}$$

$$\approx \frac{\partial \mathcal{H}(r,Q)^2_{Q=0} \langle HOMO \mid LUMO' \rangle^2}{E^{(0)} - E^{(1)}} = \frac{(-0.0032 eV)^2 * 0.28^2}{-4.4558\ eV} = -0.00000018\ eV$$

Percentage of the off-diagonal contribution to the total energy change:

$$\sum_{n\neq 0} \frac{|\langle 0|\mathcal{H}^{(1)}|n\rangle|^2}{E^{(0)} - E^{(n)}} Q^2 / \Delta E \times 100\% = -(0.00000018\ eV/0.0032\ eV) \times 100\% = 0.0056\%$$

## S-VI. VARIATION OF N–H BONDING-ENERGY

TABLE S8. ICOHP [20-22] variation between the reference ($\Delta Q = 0$, $D_{3h}$) and relaxed ($\Delta Q = 0.3889$ [u], $C_{3v}$) NH$_3$ geometries.

|  | reference structure | distorted structure | Variation of ICOHP |
|---|---|---|---|
| (N-H)$_{av}$ (eV) | -6.6383 | -6.5806 | 0.0577 |

## S-VII. MOLECULAR ORBITALS FOR THE STRUCTURE OF NH$_3$.

First-principles electronic structure calculations were performed for both the reference and distorted structures of the NH$_3$ molecule using the VASP. Based on the calculated results, the LOBSTER program [20-22,28] was employed to construct a real-space Hamiltonian within a tight-binding framework. The Hamiltonian was then diagonalized using Maple [29] to obtain the corresponding molecular orbitals, which are summarized in Tables S9 and S10. To verify our results, we repeated the electronic-structure calculations with the QE package and post-processed the output through WANNIER90 [30,31] to construct a tight-binding Hamiltonian. Diagonalizing this Hamiltonian in Maple gives a second set of molecular-orbital coefficients (Tables S11 and S12). The close agreement between the two independent sets of orbitals confirms the reliability of the computational methodology used in this work.

TABLE S9. MOs for the reference structure ($\Delta Q = 0$, $D_{3h}$) of $NH_3$. (VASP+LOBSTER+Maple)

| Molecular Orbital Compositions | Electron Population |
| --- | --- |
| $2e' = 0.35\ H1_{1s} + 0.34\ H2_{1s} - 0.65\ H3_{1s} + 0.02\ N_{2s} + 0.01\ N_{2py} + 0.01\ N_{2pz} + 0.58\ N_{2px}$ | 0 |
| $2e' = 0.61\ H1_{1s} - 0.53\ H2_{1s} + 0.06\ H3_{1s} + 0.06\ N_{2s} + 0.58\ N_{2py} - 0.01\ N_{2px}$ | 0 |
| $2a_1' = 0.39\ H1_{1s} + 0.51\ H2_{1s} + 0.48\ H3_{1s} + 0.60\ N_{2s} - 0.06\ N_{2py} - 0.02\ N_{2px}$ | 0 |
| $a_2'' = 1.00\ N_{2pz}$ | 2 |
| $e' = 0.24\ H1_{1s} + 0.24\ H2_{1s} - 0.47\ H3_{1s} - 0.82\ N_{2px}$ | 2 |
| $e' = -0.41\ H1_{1s} + 0.41\ H2_{1s} + 0.81\ N_{2py}$ | 2 |
| $a_1' = 0.35\ H1_{1s} + 0.35\ H2_{1s} + 0.35\ H3_{1s} - 0.80\ N_{2s}$ | 2 |

TABLE S10. MOs for the relaxed structure ($\Delta Q = 0.3889$ [u], $C_{3v}$) of $NH_3$. (VASP+LOBSTER+Maple)

| Molecular Orbital Compositions | Electron Population |
| --- | --- |
| $2e = 0.33\ H1_{1s} + 0.34\ H2_{1s} - 0.66\ H3_{1s} - 0.01\ N_{2pz} + 0.58\ N_{2px}$ | 0 |
| $2e = -0.55\ H1_{1s} + 0.59\ H2_{1s} + 0.03\ H3_{1s} + 0.03\ N_{2s} - 0.59\ N_{2py} - 0.02\ N_{2pz}$ | 0 |
| $3a_1 = 0.48\ H1_{1s} + 0.42\ H2_{1s} + 0.46\ H3_{1s} + 0.54\ N_{2s} + 0.03\ N_{2py} - 0.28\ N_{2pz} - 0.01\ N_{2px}$ | 0 |
| $2a_1 = 0.07\ H1_{1s} + 0.08\ H2_{1s} + 0.06\ H3_{1s} + 0.31\ N_{2s} - 0.01\ N_{2py} + 0.94\ N_{2pz}$ | 2 |
| $e = 0.24\ H1_{1s} + 0.24\ H2_{1s} - 0.48\ H3_{1s} - 0.81\ N_{2px}$ | 2 |
| $e = -0.42\ H1_{1s} + 0.42\ H2_{1s} + 0.81\ N_{2py}$ | 2 |
| $a_1 = 0.35\ H1_{1s} + 0.35\ H2_{1s} + 0.35\ H3_{1s} - 0.78\ N_{2s} + 0.18\ N_{2pz}$ | 2 |

TABLE S11. MOs for the reference structure ($\Delta Q = 0$, $D_{3h}$) of $NH_3$. (QE+WANNIER90+Maple)

| Molecular Orbital Compositions | Electron Population |
|---|---|
| $2e' = 0.59\ H1_{1s} - 0.59\ H2_{1s} + 0.55\ N_{2py}$ | 0 |
| $2e' = 0.32\ H1_{1s} + 0.32\ H2_{1s} - 0.71\ H3_{1s} + 0.03\ N_{2s} + 0.55\ N_{2px}$ | 0 |
| $2a_1' = 0.49\ H1_{1s} + 0.49\ H2_{1s} + 0.44\ H3_{1s} - 0.57\ N_{2s} + 0.03\ N_{2px}$ | 0 |
| $a_2'' = 1.00\ N_{2pz}$ | 2 |
| $e' = 0.23\ H1_{1s} + 0.22\ H2_{1s} - 0.45\ H3_{1s} - 0.84\ N_{2px}$ | 2 |
| $e' = -0.39\ H1_{1s} + 0.39\ H2_{1s} + 0.83\ N_{2py}$ | 2 |
| $a_1' = 0.33\ H1_{1s} + 0.33\ H2_{1s} + 0.33\ H3_{1s} + 0.82\ N_{2s}$ | 2 |

TABLE S12. MOs for the relaxed structure ($\Delta Q = 0.3889$ [u], $C_{3v}$) of $NH_3$. (QE+WANNIER90+Maple)

| Molecular Orbital Compositions | Electron Population |
|---|---|
| $2e = 0.59\ H1_{1s} - 0.59\ H2_{1s} + 0.55\ N_{2px}$ | 0 |
| $2e = -0.34\ H1_{1s} - 0.34\ H2_{1s} + 0.69\ H3_{1s} - 0.55\ N_{2px}$ | 0 |
| $3a_1 = 0.47\ H1_{1s} + 0.47\ H2_{1s} + 0.46\ H3_{1s} - 0.51\ N_{2s} - 0.29\ N_{2pz} + 0.01\ N_{2px}$ | 0 |
| $2a_1 = 0.08\ H1_{1s} + 0.08\ H2_{1s} + 0.08\ H3_{1s} - 0.32\ N_{2s} + 0.94\ N_{2pz}$ | 2 |
| $e = 0.22\ H1_{1s} + 0.23\ H2_{1s} - 0.45\ H3_{1s} - 0.84\ N_{2px} + 0.01\ N_{2py}$ | 2 |
| $e = -0.39\ H1_{1s} + 0.39\ H2_{1s} + 0.83\ N_{2py} + 0.01\ N_{2px}$ | 2 |
| $a_1 = 0.33\ H1_{1s} + 0.33\ H2_{1s} + 0.33\ H3_{1s} + 0.80\ N_{2s} + 0.19\ N_{2pz}$ | 2 |

# References

[1]   X. Gonze, Phys. Rev. A **52**, 1096 (1995).
[2]   S. Y. Savrasov, Phys. Rev. B **54**, 16470 (1996).
[3]   S. Baroni, P. Giannozzi, and A. Testa, Phys. Rev. Lett. **58**, 1861 (1987).
[4]   J. P. Perdew, K. Burke, and M. Ernzerhof, Phys. Rev. Lett. **77**, 3865 (1996).
[5]   G. Kresse and J. Furthmuller, Phys. Rev. B **54**, 11169 (1996).
[6]   G. Kresse and D. Joubert, Phys. Rev. B **59**, 1758 (1999).
[7]   X. Gonze *et al.*, Comput. Phys. Commun. **180**, 2582 (2009).
[8]   S. Poncé, G. Antonius, P. Boulanger, E. Cannuccia, A. Marini, M. Côté, and X. Gonze, Comput. Mater. Sci. **83**, 341 (2014).
[9]   S. Y. Savrasov and D. Y. Savrasov, Phys. Rev. B **46**, 12181 (1992).
[10]  G. Grimvall, *The electron-phonon interaction in metals.* (North-Holland Publishing Company, New York, 1981), p. 36-42.
[11]  H. Wendel and R. M. Martin, Phys. Rev. B **19**, 5251 (1979).
[12]  J. Xu, X. Huang, X. Cheng, M. H. Whangbo, and S. Deng, Angew. Chem. Int. Edit. **61**, e2021116404 (2022).
[13]  P. Hohenberg and W. Kohn, Phys. Rev. B **136**, B864 (1964).
[14]  W. Kohn and L. J. Sham, Phys. Rev. **140**, A1133 (1965).
[15]  R. O. Jones and O. Gunnarsson, Rev. Mod. Phys. **61**, 689 (1989).
[16]  J. M. Rondinelli, A. S. Eidelson, and N. A. Spaldin, Phys. Rev. B **79**, 205119 (2009).
[17]  S. Yoshida, H. Akamatsu, and K. Hayashi, Phys. Rev. Lett. **127**, 215701 (2021).
[18]  T. Nagai, Y. Mochizuki, S. Yoshida, and T. Kimura, J. Am. Chem. Soc. **145**, 8090 (2023).
[19]  P. E. Blöchl, Phys. Rev. B **50**, 17953 (1994).
[20]  R. Dronskowski and P. E. Blochl, J. Phys. Chem. **97**, 8617 (1993).
[21]  V. L. Deringer, A. L. Tchougréeff, and R. Dronskowski, J. Phys. Chem. A **115**, 5461 (2011).
[22]  S. Maintz, V. L. Deringer, A. L. Tchougréeff, and R. Dronskowski, J. Comput. Chem. **37**, 1030 (2016).
[23]  P. Giannozzi *et al.*, J. Phys. Condens. Mat. **21**, 395502 (2009).
[24]  P. Giannozzi *et al.*, J. Phys. Condens. Mat. **29**, 465901 (2017).
[25]  P. Giannozzi *et al.*, J. Chem. Phys. **152**, 154105 (2020).
[26]  D. R. Hamann, Phys. Rev. B **88**, 085117 (2013).
[27]  X. Gonze *et al.*, Comput. Phys. Commun. **205**, 106 (2016).
[28]  S. Maintz, V. L. Deringer, A. L. Tchougréeff, and R. Dronskowski, J. Comput. Chem. **34**, 2557 (2013).
[29]  M. B. Monagan, K. O. Geddes, K. M. Heal, G. Labahn, S. M. Vorkoetter, J. McCarron, and P. DeMarco, *Maple Advanced Programming Guide* (Waterloo Maple Inc, 2008).
[30]  I. Souza, N. Marzari, and D. Vanderbilt, Phys. Rev. B **65**, 035109 (2001).
[31]  G. Pizzi *et al.*, J. Phys. Condens. Mat. **32**, 165902 (2020).